\documentclass[%
 reprint,
 amsmath,amssymb,
 aps,
]{revtex4-2}

\usepackage[utf8]{inputenc} 
\usepackage{color}
\usepackage{amsmath}
\usepackage{amssymb}
\usepackage{caption}
\usepackage{subcaption}
\usepackage{graphicx}
\usepackage{physics}
\usepackage{appendix}
\usepackage{algorithm}
\usepackage{algpseudocode}
\usepackage{amsfonts}
\usepackage{verbatim}
\usepackage{float}
\usepackage{array}
\usepackage{listings}

\usepackage[unicode=true,pdfusetitle,
 bookmarks=true,bookmarksnumbered=false,bookmarksopen=false,breaklinks=true,pdfborder={0 0 0},backref=false,colorlinks=true]
 {hyperref} 
\hypersetup{colorlinks=true,citecolor=blue,linkcolor=cyan,urlcolor=blue,filecolor= green, breaklinks=true}
\usepackage{url}
\usepackage{breakurl}

\begin{document}

\title{A differentiable programming framework for spin models}

\author{Tiago S. Farias\href{https://orcid.org/0000-0002-6697-9333}{\includegraphics[scale=0.05]{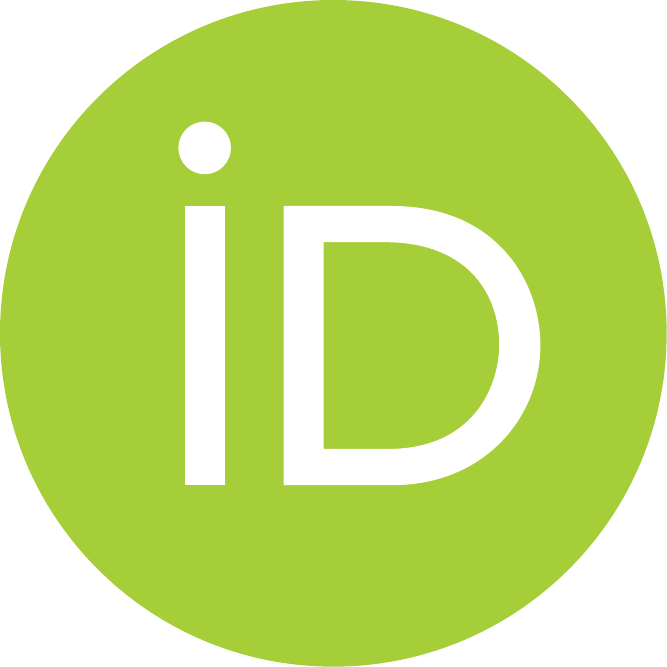}}$^{1,2,3}$}
\email[Electronic address: ]{tiago939@gmail.com}

\author{Vitor V. Schultz$^{2}$}
\email[Electronic address: ]{xultezz@gmail.com}

\author{Jos\'e C. M. Mombach$^{2}$}
\email[Electronic address: ]{jcmombach@ufsm.br}

\author{Jonas Maziero\href{https://orcid.org/0000-0002-2872-986X}{\includegraphics[scale=0.05]{orcidid.pdf}}$^{2}$}
\email[Electronic address: ]{jonas.maziero@ufsm.br}

\affiliation{$^1$ {Physics Deparament, Center for Exact Sciences and Technology, Federal University of São Carlos, Rod. Washington Luís, km 235, 13565-905, São Carlos, SP, Brazil}}
\affiliation{$^2$ Physics Department, Center for Natural and Exact Sciences, Federal University of Santa Maria, Roraima Avenue 1000, Santa Maria, RS, 97105-900, Brazil}
\affiliation{$^3$ Institute and Center for Development and Research in Software Technology - ICTS, Governor Danilo de Matos Areosa Avenue, 1199, 69075-904, Manaus, AM, Brazil}

\begin{abstract}
We introduce a novel framework for simulating spin models using differentiable programming, an approach that leverages the advancements in machine learning and computational efficiency. We focus on three distinct spin systems: the Ising model, the Potts model, and the Cellular Potts model, demonstrating the practicality and scalability of our framework in modeling these complex systems. Additionally, this framework allows for the optimization of spin models, which can adjust the parameters of a system by a defined objective function. In order to simulate these models, we adapt the Metropolis-Hastings algorithm to a differentiable programming paradigm, employing batched tensors for simulating spin lattices. This adaptation not only facilitates the integration with existing deep learning tools but also significantly enhances computational speed through parallel processing capabilities, as it can be implemented on different hardware architectures, including GPUs and TPUs.
\end{abstract}


\maketitle

\onecolumngrid

\section{Introduction}

The swift evolution in machine learning has fundamentally altered software engineering, marking significant strides in areas like computer vision \cite{computer_vision}, robotics \cite{robotics}, and protein folding \cite{alphafold}. This shift is primarily attributed to the emergence of artificial neural networks, which introduced a new programming paradigm anchored in automatic differentiation \cite{diff_prog}. At the core, neural networks employ the backpropagation algorithm \cite{backprop} that hinges on a differentiable computational graph. With automatic differentiation, the chain rule from differential calculus can be harnessed to train complex neural networks end-to-end, provided their components have well-defined derivatives.

Over the years, a multitude of frameworks have come to the forefront, including but not limited to PyTorch \cite{pytorch}, TensorFlow \cite{tensorflow}, and Jax \cite{jax2018github}. These frameworks provide comprehensive toolkits for implementing any differentiable program and are optimized to harness the power of contemporary hardware such as graphics processing units (GPUs) and tensor processing units (TPUs) \cite{tpu_paper}. Furthermore, they are also being adapted for compatibility with emerging hardware architectures, including neuromorphic computing platforms \cite{eshraghian2021training} and quantum computers \cite{pennylane}. This has fostered an ecosystem around these frameworks, enhancing their scalability across multiple devices while also boosting their computational speed and memory bandwidth.

As software and hardware continue to evolve, the landscape of programming has been reshaped by the advent of differentiable programming (DP) \cite{diffprog}. In DP, a program can be constructed by composing differentiable building blocks, allowing this paradigm to extend beyond the implementation of machine learning algorithms and impact other scientific and engineering fields, including physics simulations, optimization problems and computer graphics.

Spin models \cite{history_spins} are mathematical descriptions used to study the behavior of a system of interacting elements. Spins are representations of physical quantities, such as the orientation of magnetic moments of atoms, that can assume specific values according to the model of interest. The interaction between spins in these models is governed by a Hamiltonian associated with the energy of the system. By analyzing the statistical distribution of spins in a model, one can predict macroscopic properties, such as magnetization and specific heat. Spin models are used to study a wide variety of physical phenomena, including phase transitions \cite{miyashita_phase_2010}, cell behavior \cite{rens_energy_2019}, and neural networks \cite{kinzel_spin_1985}.

The Ising model \cite{ising} is one of the simplest spin models, consisting of only two possible spin values, usually referred to as spin up and spin down, that interact via a coupling value. This model has been extensively studied and provides the foundation for understanding the behavior of magnetic materials. Since its introduction in 1920, other models have been developed as extensions or modifications of the Ising model. One example is the Potts model \cite{ref_potts}, which differs from the Ising model by the number of values a spin can have.

Spin models have applications beyond simulating magnetic systems. Cellular models, for instance, aim to simulate the behavior of biological cells, and are derived from the formulation of spin models \cite{potts_celular}. The Cellular Potts model, also known as the Glazier-Graner-Hogeweg model, is an example that can simulate various cellular dynamics, such as morphogenesis \cite{hirashima_cellular_2017, chen_parallel_2007}, cell sorting \cite{potts_celular, durand_large-scale_2021}, and cancer spreading \cite{szabo_cellular_2013, cancer2}, making it a useful tool for studying a range of biological phenomena related to cell behavior.

In addition to leveraging the simulation of spin models for studying the magnetic characteristics of various materials, these models can be useful in addressing other complex challenges, particularly within the realm of combinatorial optimization. To utilize spin models in this manner, one can establish a map between a specific optimization problem and a corresponding spin model configuration on a lattice. This correspondence ensures that solving the given optimization problem becomes analogous to optimizing the spin lattice, subject to certain constraints. The ability to cast problems in this manner opens avenues for further expanding the utility of spin models in interdisciplinary research. Examples of applications encompass the Traveling Salesman Problem, graph theoretical analyses, and message-passing algorithms.

However, simulating spin models can be computationally expensive. They are typically simulated using Monte Carlo methods \cite{book_monte_carlo,mc}, which require many simulation steps to obtain the desired observable averages from the system. These models can also suffer from scale problems due to critical slowing down \cite{critical0, mc_simulation, gould_overcoming_1989, critical3}, resulting in low probabilities of state change at certain temperature regimes. In addition, in some models, the calculation of desired quantities can only be performed after the system has reached equilibrium, which is achieved through thermalization \cite{mc_simulation2}, whereby a certain number of Monte Carlo steps are taken before statistical values are recorded.

For most systems of interest, exact solutions to the Ising model, and other spin models, are only known for a few special cases, and numerical simulations are required to study their properties. Therefore, Monte Carlo methods are essential for simulating spin models because they enable the sampling of the space of possible configurations of the model and estimation of the thermodynamic properties of the system, which would otherwise be difficult to obtain.

In this paper, we propose using differentiable programming to simulate and optimize spin and cell models, leveraging the capabilities of DP frameworks to scale on modern hardware. The rest of this paper is organized as follows: In Sec. \ref{sec:rel_work}, we discuss related works. In Sec. \ref{sec:meth}, we present the methods we use, including the adaptations of Monte Carlo methods to the new paradigm, as well as descriptions of the systems we study in this article. In Sec. \ref{sec:resul}, we present the results obtained, and in Sec. \ref{sec:conc}, we give our final remarks.

\section{Related Work}
\label{sec:rel_work}

Differentiable programming has been applied to scientific computing tools, such as finite element methods and numerical optimization, with the aim of improving the efficiency and accuracy of these techniques. One application example  is the use of automatic differentiation to compute gradients in finite element simulations, which can be used to optimize the parameters of the simulation or to perform inverse problems. This has led to the development of several differentiable finite element libraries, such as FEniCS \cite{ScroggsEtal2022} and Firedrake \cite{rathgeber_firedrake:_2016}, which enable the efficient implementation of complex models.

Another area of interest is the integration of differentiable programming with numerical optimization techniques, such as gradient descent and conjugate gradient methods. This has been shown to be particularly useful for solving control problems \cite{dp_paper2} and inverse problems \cite{dp_inverse1, Grinis_2022, thuerey2021pbdl, hu2019difftaichi}, where the goal is to infer the parameters of a physical system from observed data. By using differentiable programming to compute gradients, it is possible to perform gradient-based optimization of these parameters, which can improve the accuracy and speed of the solution.

Recent work has also focused on the use of differentiable programming in the context of computational fluid dynamics \cite{Takahashi2021Differentiable, fan2023differentiable, dp_fluid}, where it has been shown to be effective in improving the efficiency of simulations, and can significantly reduce the computational cost of simulations while maintaining accuracy.

With respect to machine learning applied to spin models, some works propose a neural network to classify a lattice of spins by the thermodynamic phase. Ref. \cite{efthymiou_super-resolving_2019} proposes a super-resolution method to increase the size of a network without the need of simulations on large scale. Neural networks could also be used to approximate the simulation of a model. For instance, Generative Adversarial Networks can be trained to generate a sample of a lattice given a temperature \cite{liu_simulating_2017}.

It is worthwhile mentioning that many works were done on accelerating cellular and tissue modeling \cite{cpm1, cpm2, cpm3, cpm4, cpm5}. Among the Monte Carlo methods that can be used to simulate spin models, we can use Gibbs sampling \cite{gibbs}, Wolff Cluster \cite{wolff} and Metropolis-Hastings algorithm \cite{metropolis_algorithm}. This latter being the most friendly to make use of parallel computation. In the Metropolis-Hastings algorithm, applied to a spin model, a random initial state is chosen, and then the system is updated iteratively by randomly flipping one spin and calculating the change in energy. If the change in energy is negative, the new state is accepted. Else, if the change in energy is positive, the new state is accepted with a certain probability that depends on the control parameters, such as the equilibrium temperature.

\begin{figure}[t]
    \centering
    \begin{subfigure}{0.5\textwidth}
    \centering
        \includegraphics[width=0.5\linewidth]{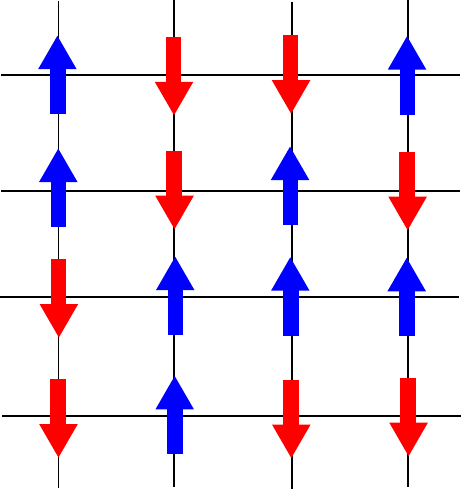}
        \caption{}
        \label{}
    \end{subfigure}%
    \begin{subfigure}{0.5\textwidth}
    \centering
        \includegraphics[width=0.5\linewidth]{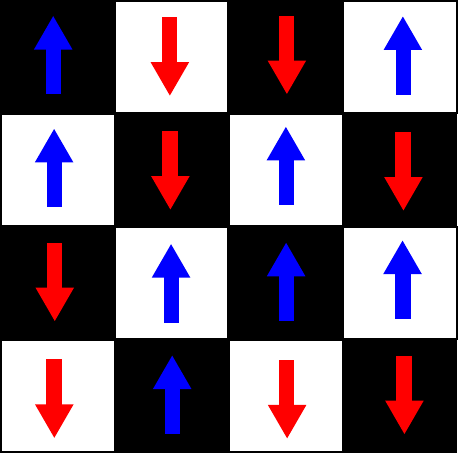}
        \caption{}
        \label{}
    \end{subfigure}
    \caption{The checkerboard method marks each spin of a lattice (a) with a color from a checkerboard pattern (b). All spins marked with the same color are updated in parallel.}
    \label{fig:checkerboard}
\end{figure}

The checkerboard method \cite{checkerboard} is a important technique that is commonly used for parallelization in the Metropolis-Hastings method. The algorithm proceeds in two steps. First, a subset of the spins is chosen, which consists of all the spins located on the black squares of a checkerboard pattern, as shown in Fig. \ref{fig:checkerboard}. The energy change resulting from a flip of each spin is calculated, and the spins are flipped with a probability given by the Metropolis algorithm. In the second step, another subset of spins is chosen, which consists of all the spins located on the white squares of the checkerboard pattern. The energy change resulting from a flip of each spin is again calculated, and the spins are flipped with a probability given by the Metropolis algorithm. The energy of the system is updated if the move is accepted. The checkerboard method is repeated for many iterations, and the spins eventually reach a state of equilibrium.

It is important to acknowledge that the order of the neighborhood in interacting spins plays a pivotal role in spin models, as it fundamentally shapes the interactions among spins. Within the framework of the original Ising model, spin interactions are confined to nearest neighbors. For these interactions, the checkerboard method is the most effective. However, the utility of this method can extend beyond this specific neighborhood arrangement. It can be applied to any level of spin interaction, provided that the central site is distinguished by one color and its surrounding neighbors by another.

\setlength{\parskip}{0pt}

The acceleration of Monte Carlo simulations for the Ising model represents a significant advancement in computational physics, offering the potential for substantial reductions in computational time. This acceleration can be achieved through the utilization of specialized hardware architectures, such as Graphics Processing Units and Tensor Processing Units, which are particularly adept at handling the parallelizable tasks inherent to these simulations. Recent studies have primarily concentrated on the development and optimization of computational kernels written in the CUDA (Compute Unified Device Architecture) language for execution on GPUs. These kernels often implement the Metropolis-Hastings algorithm in conjunction with a checkerboard updating scheme to ensure efficient state space sampling \cite{ROMERO2020107473, SPP-2023-PA-06, PREIS20094468, BLOCK20101549, WEIGEL201192, JurisicSIMULATIONOI, gpu61}. Conversely, other investigations have explored the potential for conducting cluster updates using the GPU, with the Wolff algorithm being a example of this approach \cite{KOMURA20121209}. This method represents an alternative strategy that can potentially offer improvements in efficiency and convergence rates, especially for systems near criticality.

The assessment of performance in these enhanced Monte Carlo methods typically involves measuring the number of spin flips per time unit, serving as a standard for comparison. However, it is crucial to acknowledge the susceptibility of these metrics to variation when applied across diverse hardware configurations, underscoring the importance of context in their interpretation. Table \ref{tab:gpu_papers} offers a synthesis of research employing specialized hardware to expedite Monte Carlo simulations, detailing not only the comparative performance metrics but also the capacity for algorithm optimization and the availability of the corresponding source code.

\begin{table}[t]
\begin{center}
\begin{tabular}{ | m{6em} | m{3cm}| m{2cm}| m{3cm}| m{2cm}| m{2cm}| }
  \hline
  Size & Hardware & Performance (flips/ns) & Optimization? & Code available? & Reference \\ 
  \hline
  $(20 \times 128)^2$ & Tesla V100 & 48.14 & no & yes & \cite{ROMERO2020107473} \\ 
  \hline
  $(20 \times 128)^2$ & TPU & 8.19 & no & yes & \cite{ROMERO2020107473} \\ 
  \hline
  $(20 \times 128)^2$ & RTX 3050 laptop & 0.05 & no & yes & \cite{SPP-2023-PA-06} \\
  \hline
  $(1024)^2$ & GTX 280 & 0.001 & no & no & \cite{PREIS20094468} \\ 
  \hline
  $(4096)^2$ & Tesla C1060 & 8.03 & no & no & \cite{BLOCK20101549} \\ 
  \hline
  $(16384)^2$ & GTX 480 & 0.034 & no & no & \cite{WEIGEL201192} \\ 
  \hline
  $(550)^2$ & Tesla T10 & 0.03 & no & pseudo-code & \cite{JurisicSIMULATIONOI} \\ 
  \hline
  $(4)^2$ & Tesla C1060 & 0.01 & no & no & \cite{gpu61} \\ 
  \hline
  $(4096)^2$ & GTX 285 & 0.2 & no & pseudo-code & \cite{KOMURA20121209} \\ 
  \hline
  Graph C7 & Tesla K4 & 10 seconds & yes & no & \cite{cook2019gpu} \\ 
  \hline
  \hline
  $(20 \times 128)^2$ & RTX 3060 laptop & 1.53 - 1.71 & yes & yes & \textbf{ours} \\
  \hline
  $(20 \times 128)^2$ & TPU & 0.89 - 13.57 & yes & yes & \textbf{ours} \\ 
  \hline
\end{tabular}
\end{center}
\caption{Comparative performance analysis of Monte Carlo simulations across various hardware configurations. It also details the potential for algorithm optimization and the availability of the corresponding source code. Note that performance of our work has a range which depends on the batch size.}
\label{tab:gpu_papers}
\end{table}

Optimization of complex systems can be effectively conducted through the integration of Monte Carlo methods with gradient-based analysis. This hybrid approach has demonstrated significant efficacy across a broad spectrum of applications. For instance, in the field of computer graphics, particularly with regard to rendering algorithms, Monte Carlo methods equipped with gradient information have yielded improvements in the efficiency and accuracy of image synthesis \cite{raytracing_mc}, being important to overcome the challenges associated with the stochastic nature of global illumination calculations, culminating in accelerated convergence rates and enhanced image quality.

Moreover, the calculation of expectation values, a fundamental task in probabilistic modeling and statistical analysis, has work on Monte Carlo differentiability. The incorporation of gradient information has been show to enhance the precision of Monte Carlo estimations \cite{mc_expectation}, which is useful for expediting the identification of phase transitions and enhancing the precision in ascertaining ground states of intricate two-dimensional interacting many-body systems.

\section{Methods}
\label{sec:meth}

One of the main challenges of implementing the differentiable programming paradigm is translating an algorithm in a way that could be advantageous. The Metropolis-Hastings algorithm is typically implemented using either functional or object-oriented programming. Translating this algorithm to differentiable programming requires a few modifications on how the algorithm is formulated and implemented.

In traditional programming, the Metropolis algorithm is typically implemented using a sequence of discrete steps. These steps involve updating the spin variables, computing the energy of the system, and then accepting or rejecting the proposed configuration based on an acceptance criterion. In differentiable programming frameworks, we can express an array of elements as a batched tensor with sizes up to five dimensions. For instance, deep learning applied in computer vision usually uses an array of images that can be represented as a tensor with size $[B,C,H,W]$, with $B$ the batch dimension, $C$ the channel dimension and $H,W$ the height and width of the images respectively. 

Since the 4-dimensional format of $[B,C,H,W]$ is compatible with many modern deep learning frameworks, this array is useful for appling deep learning techniques to two-dimensional spin lattices. This can be particularly advantageous when using differentiable programming to simulate spin models, as it allows for seamless integration with existing deep learning tools and techniques. Additionally, the use of a batch dimension allows faster processing of multiple spin lattices simultaneously. This can be useful when simulating large-scale spin models, as it enables parallel processing of multiple samples or multiple temperatures at the same time.

\subsection{Ising model}

The Ising model consists of two states, called spins, which physically represent the magnetic moment of materials. They can be in a up state $(\sigma=+1)$ or down state $(\sigma=-1)$. This model has a phase transition in certain lattice geometries, where a change on the behavior of physical quantities, such as the collective magnetic moment, occurs. For example, on a 2D square lattice with $J < 0$, the Ising model predicts a change from a paramagnetic phase, characterized by a random mixture of spins, to a ferromagnetic phase, characterized by a alignment of the spins. The Hamiltonian that describes the energy of the system is:

\begin{equation}\label{equ:ising_model}
    \mathcal{H} = \sum_{i,j} J_{ij} \sigma_i \sigma_j + \sum_i B_i \sigma_i,
\end{equation}
with $J_{i,j}$ being the interaction strength between the spins $\sigma_i$ and $\sigma_j$, and $B_i$ an external magnetic field on the spin $\sigma_i$. 

The modified Monte Carlo simulation of the Ising model using the Metropolis-Hastings algorithm requires a convolution operation to calculate the system's Hamiltonian. This convolution depends on two topological conditions of the system: its dimension and connectivity. The dimension of the system directly determines the dimension of the convolution, with a 1D convolution used for one-dimensional spin networks, 2D convolution for square or triangular networks, and 3D convolution for cubic networks, and so on. The connectivity of the system, which describes how the sites are connected, determines the shape of the convolution kernel.

For example, consider a square network with first-neighbor interactions. Each site is connected to its four nearest neighbors, which are located above, below, to the right and to the left of it. In this case, the convolution kernel has size of $3\times3$, with values of $1$ in the positions corresponding to the neighboring sites and $0$ everywhere else, including the center (to prevent self-interaction of the spins).

By using this convolution, it becomes possible to 
simulate the behavior of the Ising model with DP frameworks.
The convolution operation provides an 
practical way to calculate the Hamiltonian of a system, which is a crucial step in the Metropolis-Hastings algorithm.

The kernel of the convolution depends on the geometry of the lattice. For the square lattice, the kernel is:
\begin{equation}
    K = \begin{bmatrix}
0 & 1 & 0\\
1 & 0 & 1 \\
0 & 1 & 0
\end{bmatrix}
\end{equation}
Thus, for each spin the energy is obtained by the top, down, left and right neighbors, which corresponds to the square lattice interaction. Note that the kernel shape does not need to be square, as long as it accounts for the geometric shape of the network. The shape of the kernel is important because it determines the specific features that are extracted from the network. For instance, if the kernel is square, it may extract different features compared to a triangular or hexagonal kernel. However, as long as the kernel accounts for the geometric shape of the network, it can be any shape that is suitable for the particular analysis.

For lattices with other connectivity, such as the triangular lattice, a transformation is necessary to convert the triangular lattice into a square lattice so that a square kernel can be used. This transformation involves adding null spins to the left and right of the central spin, which effectively creates a rectangular shape for the kernel:
\begin{align}
K = \begin{bmatrix}
0 & 1 & 0 & 1 & 0\\
1 & 0 & 0 & 0 & 1 \\
0 & 1 & 0 & 1 & 0
\end{bmatrix},\hspace{10pt}
&
K = \begin{bmatrix}
1 & 0 & 1\\
0 & 0 & 0 \\
0 & 1 & 0
\end{bmatrix}.
\end{align}

By applying the convolution operation to the lattice using these rectangular kernels, the algorithm produces a map that is associated with the sum of the neighbor spin values  for each site. This map can be used then to obtain the Hamiltonian of each site.

The output of the convolution operation applied to the network is a map that is associated with the sum of the first neighbor spins. This map provides information about the local interactions between spins in the lattice. By multiplying this map produced by the convolution operation with the spin network itself, the Hamiltonian of each site in the lattice can be obtained, as described in Algorithm \ref{algorithm1}.

One requisite to optimize a spin lattice is that the algorithm must be fully differentiable in order to update the parameters. The standard Metropolis algorithm has one non-differentiable step: the update state that depends on a random number. To circumvent this problem, we replace this step by a logistic function, which is differentiable. This function has a hyper-parameter $\alpha$, which at the limit $\alpha \rightarrow \infty$ becomes the standard Metropolis algorithm.

\begin{minipage}{0.46\textwidth}
\begin{algorithm}[H]
    \centering
    \caption{Standard Metropolis-Hastings algorithm with parallelization}\label{algorithm1}
    \begin{algorithmic}[1]
        \State  Propose new random states for the lattice;

        \State Apply 2D convolution with kernel selected from the model and connectivity in both states;
    
        \State Multiply the result of the convolution to its respective state;
    
        \State Apply the checkerboard algorithm;
        
        \State Evaluate the variation in energy $\Delta E$ for each site with same checkerboard color;
        
        \If{$\Delta E \leq 0$} 
            \State Accept the change
        \Else
            \State Accept the change with probability $p=e^{-\beta\Delta\mathcal{H}}$
        \EndIf 
    \end{algorithmic}
\end{algorithm}
\end{minipage}
\hfill
\begin{minipage}{0.46\textwidth}
\begin{algorithm}[H]
    \centering
    \caption{Differentiable Monte Carlo}\label{algorithm2}
    \begin{algorithmic}[1]
        \State  Propose new random states $\mathbf{s'}$ for the lattice;

        \State Apply 2D convolution with kernel selected from the model and connectivity in both states;
    
        \State Multiply the result of the convolution to its respective state;
    
        \State Apply the checkerboard algorithm;
        
        \State Evaluate the variation in energy $\Delta E$ for each site with the same checkerboard color;

        \State Evaluate the probability of transition $p = e^{-\beta \Delta E}$;
        \State Generate random numbers \textbf{r};

        \State Calculate $q = (1 + \exp(-\alpha(p-r)))^{-1}$;

        \State $s \leftarrow qs' + (1-q)s$;

        \State Optimize an objective function $\mathcal{L}$;

        \State Discretize the spin values.
    \end{algorithmic}
\end{algorithm}
\end{minipage}

Note that the algorithm produces intermediary states because the differentiability condition $s \leftarrow q s' + (1-q)s$, which is important for the optimization of an objective function. In the case of the Ising model, we can have spin values between $-1$ and $+1$. The last step corrects by discretizing the values after updating the parameters. This can be done, in the case of the Ising model, by applying the step function:
\begin{equation}
    s \leftarrow
    \begin{cases}
      -1 \mbox{ if } s < 0\\
      +1 \mbox{ if } s \ge 0
    \end{cases}\,.
\end{equation}

Algorithm \ref{algorithm2} delineates the comprehensive modifications required for optimizing spin models. It is important to emphasize that these adjustments are crucial only in the context of an optimization problem. In scenarios where optimization is not a factor, the original algorithm presents a more appropriate choice. This distinction highlights the tailored application of Algorithm 2, ensuring its deployment is aligned with the specific requirements of the optimization task at hand.

\subsection{Potts model}

The Potts model \cite{potts_model} is a lattice model that describes the behavior of a system of interacting spins, which can take on more than two possible states. Unlike the Ising model, which has spins that can only take on two states (up or down), the Potts model allows spins to take on $q$ different states, with $q$ being any integer greater than or equal to $2$. Each spin is represented by an integer variable $\sigma_i$ that can take on values from 0 to $q-1$.

The Hamiltonian for the Potts model is given by 
\begin{equation}\label{equ:potts_model}
    \mathcal{H} = \sum_{i,j} J_{i,j} \delta (\sigma_i, \sigma_j) + \sum_i B_i \sigma_i,
\end{equation}
where the first term represents the interactions between neighboring spins and the second term represents the effect of an external magnetic field on each spin. The coupling between spins $J_{i,j}$ is a constant that depends on the interaction between spins $i$ and $j$. The Kronecker delta function $\delta (\sigma_i, \sigma_j)$ equals $1$ if $\sigma_i=\sigma_j$ and $0$ otherwise.

The Potts model has applications in various fields, such as statistical physics, materials science, and computer science. It can be used to model phase transitions \cite{potts_phase}, magnetic ordering \cite{potts_ordering}, and coloring of graphs \cite{potts_graph}. The model has also been used in image processing and computer vision, where it can be used to cluster pixels based on their colors or textures \cite{potts_image}.

To simulate the Potts model, the Metropolis-Hastings algorithm can be used, similarly to what is done for the Ising model. The simulation involves selecting a random spin and attempting to change its state to a new value using a trial move. If the energy change resulting from the trial move is negative, the move is accepted. If the energy change is positive, the move is accepted with a probability given by a acceptance probability. In the Potts model, a spin flip is not well-defined since the spin can take on more than two states. Instead, a random spin is chosen to undergo a state change, with the new state being chosen from the $q-1$ possible values that are different from the current state.

The differentiable programming Metropolis-Hastings algorithm in the Potts model follows a structure similar to that used for the Ising model, by utilizing convolution. However, the main difference between the two models lies in the Potts model's convolution, which employs more than one convolutional filter, with each filter having its own kernel. The number of kernels is determined by the geometric properties of the system of interest, and each kernel accounts for a central site neighbor.

For instance, in the case of a square lattice with first neighbor interaction, four kernels are required, as shown below:
\begin{align}\label{equ:kernel_potts}
K_1 = \begin{bmatrix}
0 & 1 & 0\\
0 & 0 & 0 \\
0 & 0 & 0
\end{bmatrix},\ 
K_2 = \begin{bmatrix}
0 & 0 & 0\\
1 & 0 & 0 \\
0 & 0 & 0
\end{bmatrix}, \
K_3 = \begin{bmatrix}
0 & 0 & 0\\
0 & 0 & 1 \\
0 & 0 & 0
\end{bmatrix},\ 
K_4 = \begin{bmatrix}
0 & 0 & 0\\
0 & 0 & 0 \\
0 & 1 & 0
\end{bmatrix}.
\end{align}

The separation of the kernels into four distinct filters is due to the Kronecker delta, which requires that each interaction with a neighbor to be accounted for separately. This contrasts with the Ising model, where the sum of neighboring spins multiplied by the central spin is sufficient.

After applying convolution with the four filters, four maps are generated for each site. To obtain the Hamiltonian of the system, the difference between each map and the spin configuration is computed. If the spins are equal, the value of the map at that position is set to zero; otherwise, the value 1 is assigned. The resulting four maps are summed to generate a single map, which, multiplied by the spin configuration of the lattice, represents the Hamiltonian of the system, which can then be used to compute various variables of interest.

\subsection{Cellular Potts model}


Cells arrange themselves spatially during morphogenesis, wound healing, or regeneration to build new tissues or restore damaged ones.
There are a number of intercellular methods of interaction that are used to carry out these functions.
Cell-cell adhesion is a key process that can direct cell motion in a manner similar to the separation of immiscible liquids due to interfacial tensions, where the liquid phases are associated with various cell types (e.g. cells of retina, liver, etc).  A well known phenomenon is the spontaneous separation of randomly mixed cell types which is known as cell sorting \cite{DAH}. 
The Cellular Potts Model (CPM), developed by Graner and Glazier, takes into account cellular adhesion to explain cell sorting and is frequently used to describe cellular processes \cite{gg}. 

In the two dimensional CPM, cells are represented as identical spin domains on a lattice. Each spin $\sigma$ has a position $\vec{x}$ and a type index $\tau \left( \sigma \right)$ (a second quantum number). The Hamiltonian (\ref{h}), which represents cell-cell adhesion and interaction with cell culture media  (a source of nutrients), describes the dynamics.  Cell area is controlled by a second quadratic energy constraint. The medium around the cell aggregate is represented mathematically as a (big) cell with unconstrained area.  The Hamiltonian is written as follows
\begin{equation}
\mathcal{H}=\frac{1}{2}\sum_{\vec x}\sum_{\vec y}^{N_{\max}} J_{ \tau(\sigma(\vec x)), \tau(\sigma(\vec y)) }
\left(1-\delta_{ \sigma(\vec x), \sigma(\vec y) }\right)
+ \lambda\sum_{\sigma}(a_{\sigma}-A_{\sigma})^2 \,\,\,,
\label{h}
\end{equation}
where $J_{ \tau(\sigma(\vec x)), \tau(\sigma(\vec y)) }$ are the cell-cell or cell-medium adhesion energies that depend on the cell type $\tau$, $\delta$ is the Kronecker's delta function, $\lambda$ is a Lagrange multiplier that controls cell area compressibility, $a_{\sigma}$ is the cell area and $A_{\sigma}$ is the cell target area. By specifying energy only at the cell-cell contact, the Hamiltonian's first term in Eq. (\ref{h}) mimics cellular adhesion. 

The CPM simulates cell motion driven by the cytoskeleton by performing
attempts to change the value in a given lattice position for another in its vicinity,
what causes domain boundary motion representing cellular membrane motion.
By selecting a target spin ($\sigma(\vec x)$) and a neighbor spin ($\sigma(\vec y)$) at random from the cell lattice, the system dynamic occurs.
The change in the target spin value for its neighbor's spin value is then decided utilizing the aforementioned algorithm 1. 

When implementing the cellular Potts model, using simple convolutions like in previous models is not sufficient because the interaction between spins, denoted by $J$, is no longer constant. Instead, it depends on the spin value and the type of cell. To address this issue, other differentiable programming techniques are used. One operation that is particularly useful for implementing convolutions is called unfolding \cite{book_unfold}.

Unfolding involves dividing a set of elements, such as a vector, matrix, or other multidimensional set, into smaller, equally sized partitions called ``patches''. The size of these patches depends on the parameters of the convolution, such as the stride and padding. For example, if we apply unfolding with padding 0, stride 4, and a kernel of size $4\times4$ to an $8\times8$ pixel image, we would obtain four patches of the image, each with a size of $4\times4$ pixels.

The cellular Potts model involves interactions between neighboring cells, and the properties of the unfolding depend on the nature of these interactions. Typically, an odd number is used to account for the central cell. For instance, if first-neighbor interactions are considered, the kernel size will be $3\times D$, where $D$ is the dimension of the system. The stride will be set to 1 to compute the energy value for each cell, while the padding will depend on the boundary conditions specified, just as in the Ising and Potts models. If second-neighbor interactions are considered, the kernel will be larger, with a size of $5\times5$ to account for more distant cells.

Once the unfolding and padding operations are applied, the next step is to copy the central spin value of each patch to the same size as the patches. This allows us to obtain a consistent representation of the spin values across the entire grid. Finally, once the central sites have been copied, each patch is compared to its corresponding copy of the central sites, element by element. During this comparison, the interaction values are assigned based on the spin value and cell type. After the comparison, the energy map per site of the system is generated by summing the values of each compared patch.








\section{Results}
\label{sec:resul}

The differentiable programming Metropolis-Hastings algorithm was employed to simulate all models discussed above. The simulations were carried out using a computing system comprising an Intel Core i7-11800H CPU, a Nvidia GeForce RTX 3060 GPU, and a TPU which was utilized through Google Colab \cite{colab}. We used the Python programming language and the PyTorch framework for deep learning. For the Ising and Potts models, we simulate lattices with different sizes, with $J=-1$ and first-order neighbor interactions. 

The simulations of the spin models were separated into two distinct operational modes: inference and training. The inference mode focused exclusively on simulating the spin interactions. Conversely, the training mode, applied to the Ising model, involved the optimization of parameters within a predefined objective function. 

The cellular Potts model considers the simulation of cell behavior. In this simulation, three cell types were chosen: light cells, dark cells, and the medium, following the original work of 1993 \cite{gg}. Additionally, 37 different spin values were chosen, with a value of 0 representing the medium, where each spin value leads to the formation of a unique cell. The chosen interaction energies are presented in Table \ref{tab:cpm_values}, and these interactions are symmetric (the interaction value is the same if site A interacts with site B or B with A). The interaction uses the Moore neighborhood with second neighbors. For the other parameters of the models, the temperature was set to $k_B T = 8$, target volume $A_{\sigma}=25$ and cell area compressibility $\lambda=1$.

\begin{table}[t]
\begin{center}
\begin{tabular}{ | m{8em} | m{1cm}| }
  \hline
  Interaction & $J_{ \tau, \tau^{\prime} }$ \\ 
  \hline
  Medium-Medium & $0$ \\ 
  \hline
  Medium-Dark & $16$ \\ 
  \hline
  Medium-Light & $16$\\
  \hline
  Light-Light & $2$\\ 
  \hline
  Dark-Dark & $14$\\ 
  \hline
  Light-Dark & $11$\\ 
  \hline
\end{tabular}
\end{center}
\caption{The interaction values $J$ depend on the cell type, for the cellular Potts model.}\label{tab:cpm_values}
\end{table}

\subsection{Ising model}

\begin{figure}[bt]
    \centering
    \begin{subfigure}{0.275\textwidth}
        \includegraphics[width=0.8\linewidth]{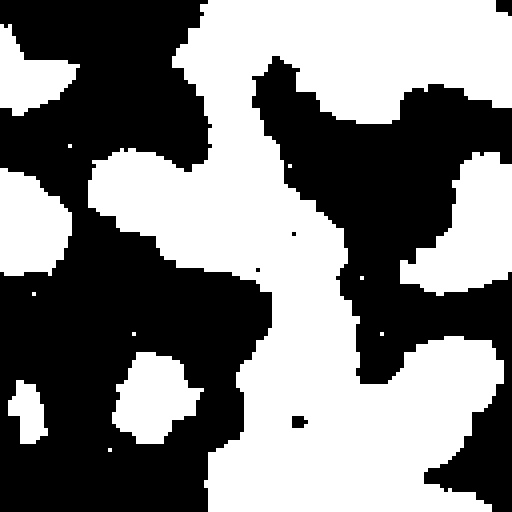}
        \caption{}
        \label{}
    \end{subfigure}%
    \begin{subfigure}{0.275\textwidth}
        \includegraphics[width=0.8\linewidth]{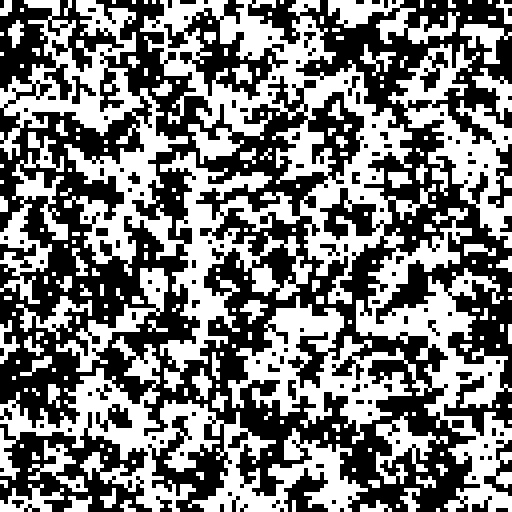}
        \caption{}
        \label{}
    \end{subfigure}
    \begin{subfigure}{0.44\textwidth}
        \includegraphics[width=1\linewidth]{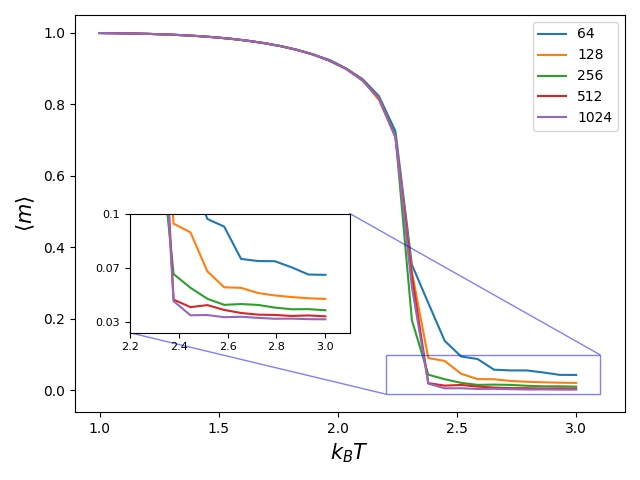}
        \caption{}
        \label{}
    \end{subfigure}
    \caption{Lattice configurations for two different temperatures of the Ising model. (a) $T < Tc$ and (b) $T > Tc$. (c) Magnetization with respect of the temperature for different lattice sizes. The temperature is measured in units of the Boltzmann constant $k_B$.}
    \label{fig:ising_simulation}
\end{figure}

Figure \ref{fig:ising_simulation} shows snapshots of the square lattice Ising model for two different temperatures and the average magnetization for different lattice sizes. The temperature of the system determines the probability of the spins flipping between up and down states. At low temperatures, the spins tend to align with their neighbors, forming large clusters that grow in size as the temperature decreases. This is due to the reduction in thermal energy, which favors the alignment of neighboring spins. These large clusters are known as domains, and they play an important role in the behavior of the system.

As the critical temperature is approached, the influence of long-range correlations between spins becomes more pronounced, leading to a change in the collective behavior of the system. At this temperature, the system undergoes a phase transition, characterized by the emergence of long-range correlations and critical fluctuations. This is a critical point where the properties of the system change abruptly. As the size of the lattice increases, the critical temperature converges towards a single value.

Above the critical temperature, at high temperatures, the domains disappear, and the system becomes disordered, with no long-range correlations. Thermal fluctuations dominate the behavior of the lattice, leading to random changes in the spin values. As the temperature increases, the magnetic spins become increasingly disordered, leading to a loss of magnetization in the system.


\begin{figure}[t]
     \centering
        \includegraphics[scale=0.75]{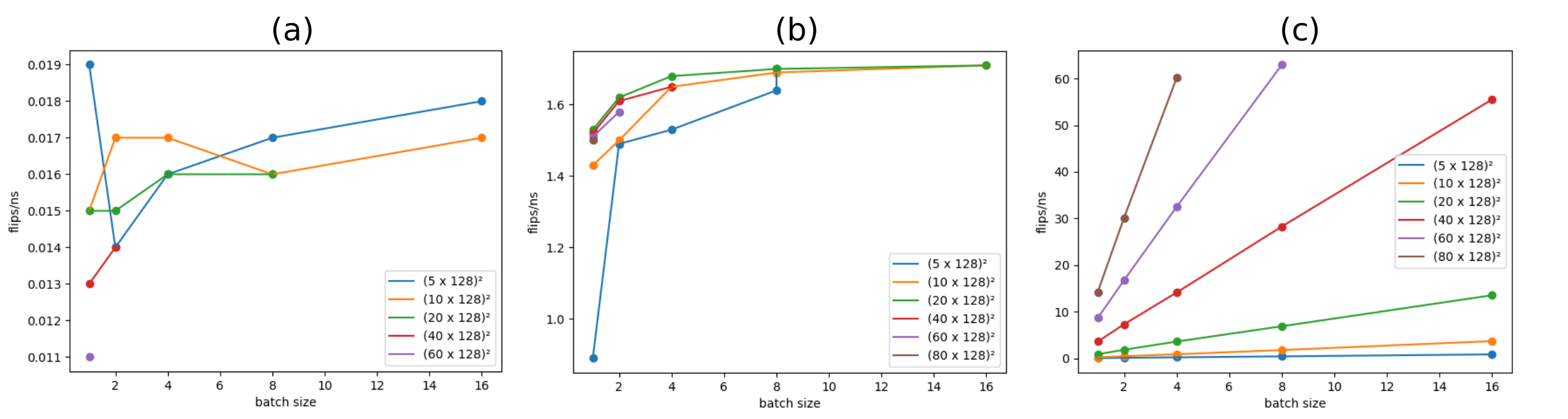}
        \caption{Flips per nanosecond for different lattice and batch sizes. (a) Simulation on CPU, (b) GPU and (c) TPU.}
        \label{fig:time_ising}
\end{figure}

The results for three different hardware are presented in Figure \ref{fig:time_ising}. Simulations run on CPU show consistent runtimes across varying batch and lattice sizes. On the other hand, GPU simulations outperform CPU by nearly 100 times in terms of speed. Notably, TPU simulations demonstrate a significant advantage in runtime as the lattice size and batch size increases, with a speedup of 10x compared to GPU simulations.

Table \ref{tab:ising_flips} presents the main results derived from Figure \ref{fig:time_ising}, encapsulating both the minimal and maximal lattice and batch sizes alongside their corresponding performance metrics across the three hardware configurations. Notably, the CPU exhibited the least effective performance in comparison to its counterparts. However, there is an observed relationship between the lattice and batch sizes, which directly influenced the performance efficiency. Specifically, it was found that both GPU or TPU demonstrated superior efficiency at different scales, measured in terms of the highest flips per nanosecond (flips/ns), contingent upon these size parameters.

\begin{table}[t]
    \begin{minipage}{.32\linewidth}
      \caption*{}
      \centering
        \begin{tabular}{| m{2cm} | m{1cm} | m{2cm} |}
            \hline
            Lattice size & Batch size & Performance (flips/ns)\\
            \hline
            $(5 \times 128)^2$ & 1 & 0.019\\
            \hline
            $(5 \times 128)^2$ & 16 & 0.018\\
            \hline
            $(20 \times 128)^2$ & 1 & 0.013\\
            \hline
            $(20 \times 128)^2$ & 8 & 0.016\\
            \hline
        \end{tabular}
    \end{minipage}%
    \begin{minipage}{.32\linewidth}
      \centering
        \caption*{}
        \begin{tabular}{| m{2cm} | m{1cm} | m{2cm} |}
            \hline
            Lattice size & Batch size & Performance (flips/ns)\\
            \hline
            $(5 \times 128)^2$ & 1 & 0.89\\
            \hline
            $(5 \times 128)^2$ & 16 & 1.70\\
            \hline
            $(20 \times 128)^2$ & 1 & 1.53\\
            \hline
            $(20 \times 128)^2$ & 16 & 1.71\\
            \hline
        \end{tabular}
    \end{minipage} 
    \begin{minipage}{.32\linewidth}
      \centering
        \caption*{}
        \begin{tabular}{| m{2cm} | m{1cm} | m{2cm} |}
            \hline
            Lattice size & Batch size & Performance (flips/ns)\\
            \hline
            $(5 \times 128)^2$ & 1 & 0.06\\
            \hline
            $(5 \times 128)^2$ & 16 & 0.86\\
            \hline
            $(20 \times 128)^2$ & 1 & 0.89\\
            \hline
            $(20 \times 128)^2$ & 16 & 55.52\\
            \hline
        \end{tabular}
    \end{minipage} 
    \caption{Performance, measured in flips/ns, for lowest and highest lattice and batch sizes. CPU (table on the left) has lowest performance, while there is a trade-off between GPU (central table) and TPU (table on the right) at different sizes.}
    \label{tab:ising_flips}
\end{table}

\subsection{Potts model}

\begin{figure}[H]
     \centering
     \includegraphics[scale=0.75]{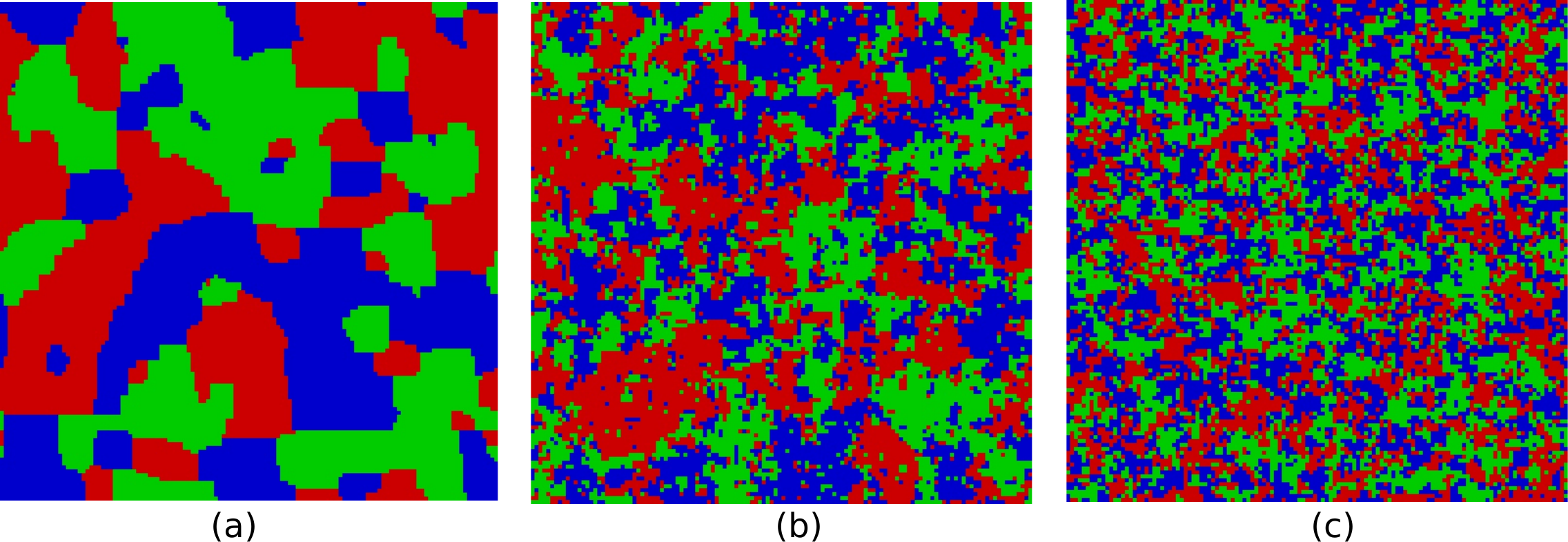}
        \caption{Lattice configurations for three different temperature of the Potts model with $q=3$. Each color represents a spin state. (a) $T < Tc$, (b) $T \approx Tc$, (c) $T > Tc$.}
        \label{fig:simulation_potts}
\end{figure}

In the case of the Potts model, the spins can take on several different states, resulting in a more complex system with a richer set of behaviors. As the temperature is lowered, the spins tend to align themselves to form distinct clusters, as shown in Fig. \ref{fig:simulation_potts}. This behavior is reminiscent of ferromagnetism, in which the magnetic moments of individual atoms align themselves in the same direction.

On the other hand, at higher temperatures, the spins in the Potts model become more disordered and exhibit more frequent state changes, resulting in a more random and unpredictable system. This behavior is similar to what is observed in the Ising model, where, at high temperatures, the magnetic moments of individual atoms become more disordered and fluctuate rapidly.

The Potts model has the same computational complexity as the Ising model. Thus the time required to flip the spins is similar for both models. This is because flipping the spin of a single site in the Potts model requires the calculation of the energy difference between the current and proposed spin states, just as in the Ising model. However, in the Potts model the energy difference depends on the number of neighboring spins that have different states, which leads to a more complex calculation than in the Ising model.

Despite this additional complexity, the computational cost of flipping the spins in the Potts model is still comparable to that of the Ising model. The number of neighboring spins is typically small compared to the total number of spins in the system. As a result, simulations of the Potts model can be performed with similar computational resources and time requirements as those for the Ising model.

\subsection{Cellular Potts model}

\begin{figure}[H]
     \centering
     \includegraphics[scale=0.75]{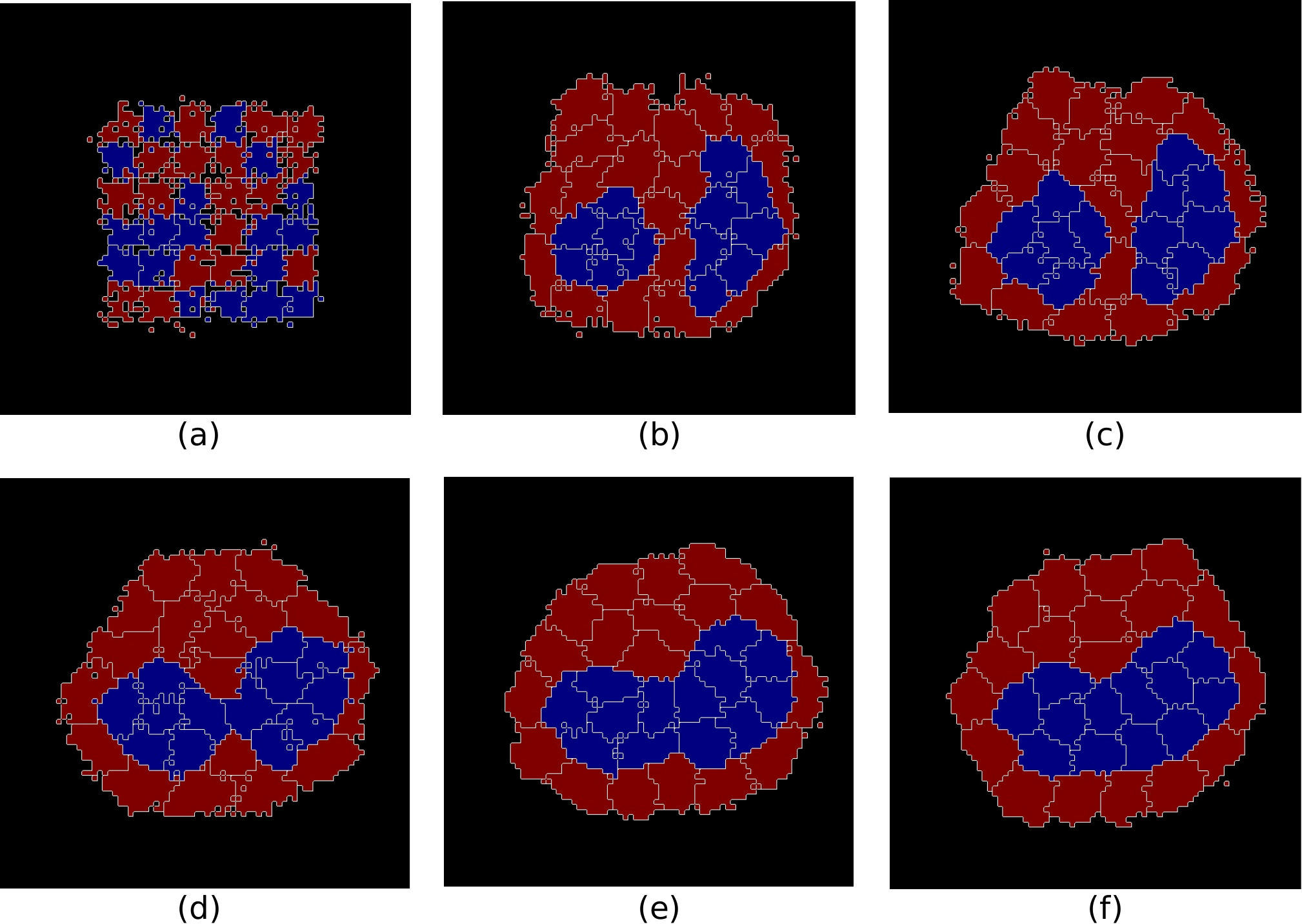}
     \caption{Cellular Potts Model simulation. Each figure is a snapshot from different steps of the simulation. The cells spontaneously reorder themselves into clusters of same type as the the number of Monte Carlo steps increases.}
        \label{fig:simulation_cpm}
\end{figure}

The evolution of a cell aggregate can be observed in Figure \ref{fig:simulation_cpm}, where snapshots of the system's states are shown with increasing the number of Monte Carlo steps. Starting from an arbitrary aggregate of square-shaped cells, the system undergoes a process that aims to minimize the number of energy-costly boundaries, resulting in the sorting of the two cell types.

\begin{figure}[H]
     \centering
      \includegraphics[scale=0.75]{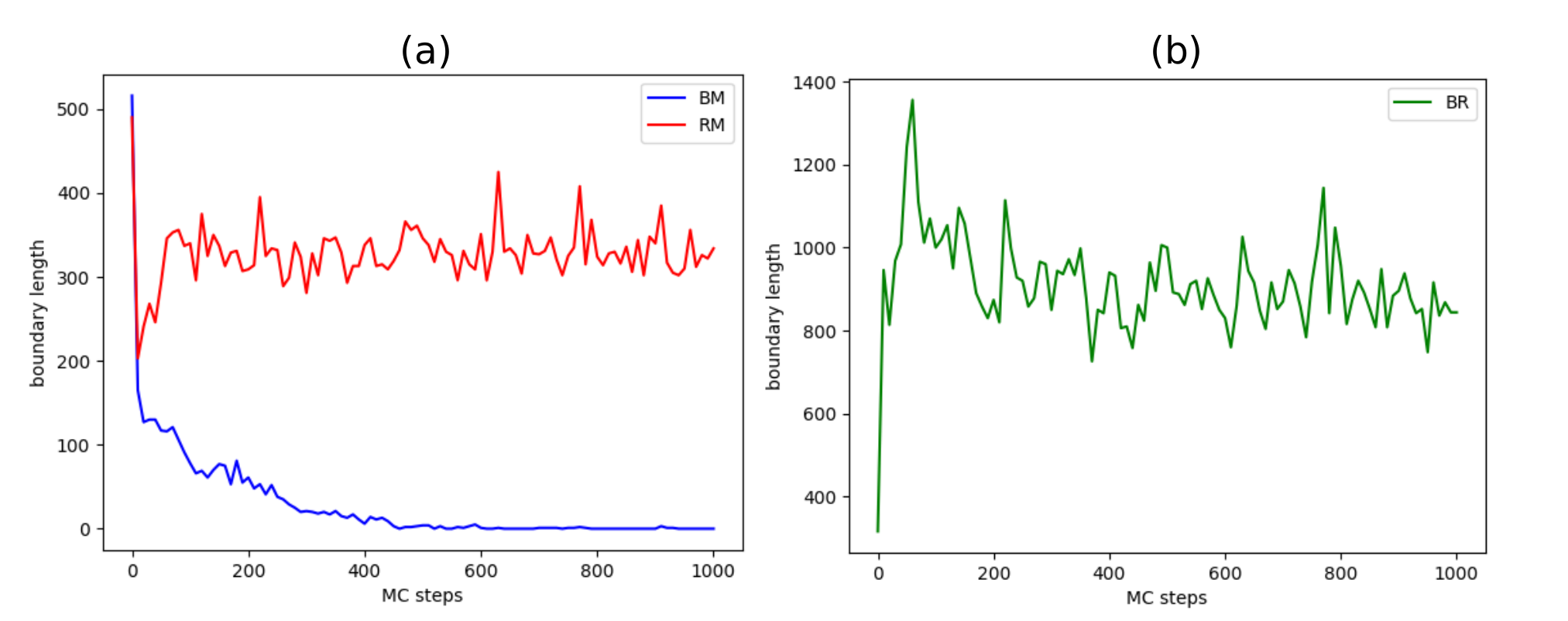}
     \caption{Boundary length between different cells types. (a) Blue cells and medium (blue line) and red cells and medium (red line). (b) Blue and red cells: we observe that the boundary length between blue cells and the medium goes to zero as the red cells surround the blue cells.}
        \label{fig:cpm_boundary}
\end{figure}

The boundary length between each cell type can be observed in Fig. \ref{fig:cpm_boundary}, showing the evolution of the system. As the simulation progresses, we observe the interface between the blue cells and the medium vanishing, while the boundary length between red cells with medium and red cells with blue cells approaches a minimum value.

The simulation's results suggest that the energy constraints in the system drive the behavior of the cell aggregate towards a more stable configuration, where the number of energy-costly boundaries is minimized. The decreasing boundary length between the different cell types indicates that the cells are actively interacting with each other, eventually sorting themselves into more cohesive groups. 

The phenomenon of cell sorting is a well-known example of self-organization in biological systems. It has been studied extensively both experimentally and theoretically, and its underlying mechanisms are thought to involve a complex interplay of various physical and biochemical processes \cite{durand_large-scale_2021, nakajima_kinetics_2011}.

One of the key factors that influence cell sorting is the interaction between cells and their surrounding environment. Some studies have shown that the CPM can be used to model not only cell sorting but also other cellular behaviors such as movement \cite{guisoni_modeling_2018} and migration \cite{axioms10010032}. By varying the values of interaction $J$ between cells, it is possible to simulate different scenarios and study their effects on cellular behavior. Since the difference among these different phenomena are only in the interaction values, the computational cost of simulating the model with differentiable programming remains the same.

\subsection{Ising optimization}

In this section, we demonstrate a proof of concept for Differentiable Monte Carlo (DMC), showcasing its application to two distinct optimization challenges: firstly, the optimization of exchange parameters $J$ between spins, and secondly, the optimization of the spin values themselves.

\subsubsection{Optimization of the exchange parameters}

State preparation is a important problem for many fields. It is known in quantum computing that general state preparation is an exponentially difficult problem \cite{gao_efficient_2017, ciavarella_state_2023, atanasova_stochastic_2023} Here we show how to optimize a spin lattice with the differentiable Monte Carlo algorithm in order to do state preparation. This differs from other state preparations techniques \cite{boltzmann_gen} that our method tries to solve this problem with a Markov chain. The parameters to be optimized are the exchange interaction  between spins $J_{ij}$, with no external magnetic field. We define the loss function as the difference between a target spin configuration $\hat{x}$ and the current spin configuration $x$, defined as the norm $L_2$:
\begin{equation}
    \mathcal{L} = ||\hat{x} - x||_2^2.
\end{equation}

It is important to note that the size of the gradients depends on the temperature of the system. At high temperatures, thermal fluctuations dominate the spin flips, preventing a spin to reaching a specific state. At low temperatures, the system may take a long time to flip. Thus, there is a intermediary temperature where changing the parameters $J_{ij}$ is optimal. This reflects on the size of the gradients during training. 

\begin{figure}[t]
    \centering
    \begin{subfigure}{0.5\textwidth}
    \centering
        \includegraphics[width=0.85\linewidth]{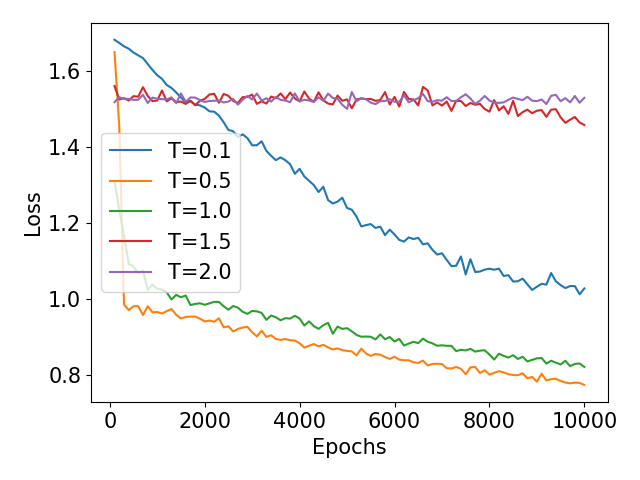}
        \caption{}
        \label{fig:optimizationT:a}
    \end{subfigure}%
    \begin{subfigure}{0.5\textwidth}
    \centering
        \includegraphics[width=0.85\linewidth]{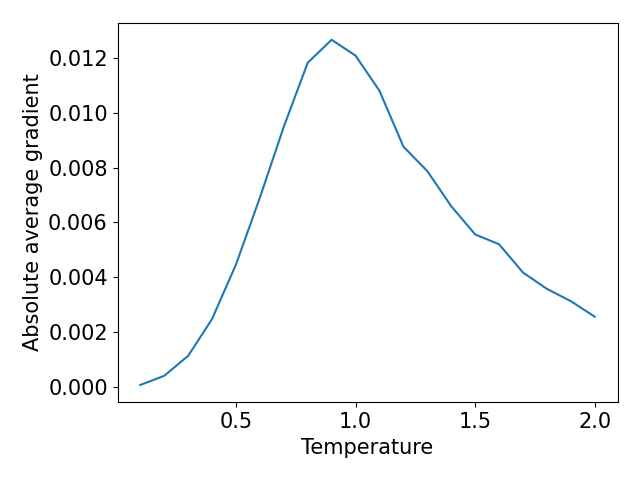}
        \caption{}
        \label{fig:optimizationT:b}
    \end{subfigure}
\caption{The success of the optimization will depend on the temperature. (a) High and low temperature prevents the training of the model. (b) There is an optimal temperature (here shown for $T=0.9$) where the average size of the gradients is maximum.}
\label{fig:optimizationT}
\end{figure}

Figure \ref{fig:optimizationT} shows how the gradient (the derivative of the loss with respect to the parameters $J_{ij}$) changes with the temperature. While the loss function does not depend explicitly on the temperature, the optimization uses Monte Carlo, which changes the spins depending on the temperature. We can see that very high and low temperatures will make the loss decrease slower compared to the optimal intermediary temperature. This behavior happens due to the size of the gradient, which depends on the temperature, as shown in Figure \ref{fig:optimizationT:b}. For large temperatures, the spins have a high energy in which they change spin randomly, which affects the optimization. Also, for low temperatures, the chance of changing a spin with MC is low, which makes the optimization also difficult. Higher gradients will change the parameters closer to its optimal solution. 

During the optimization, we used the Adam optimizer, with learning rate $\eta=0.001$, $\beta_1=0.9$ and $\beta_2=0.999$. Once we defined the hyper-parameters, we can train the system to optimize some objective function. Here we choose three different patterns that we want a spin system to achieve.

\begin{figure}[H]
    \centering
    \begin{subfigure}{0.19\textwidth}
        \includegraphics[width=0.5\linewidth]{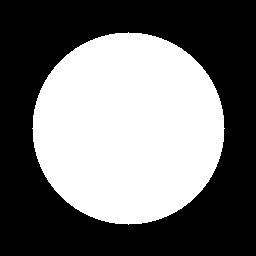}
        \caption{}
        \label{}
    \end{subfigure}%
    \begin{subfigure}{0.19\textwidth}
        \includegraphics[width=0.5\linewidth]{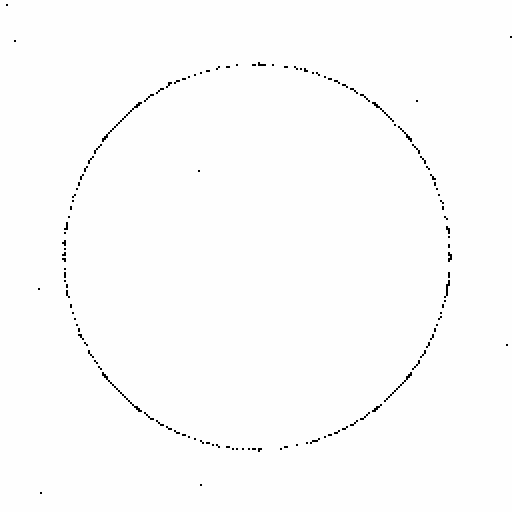}
        \caption{}
        \label{}
    \end{subfigure}
    \begin{subfigure}{0.19\textwidth}
        \includegraphics[width=0.5\linewidth]{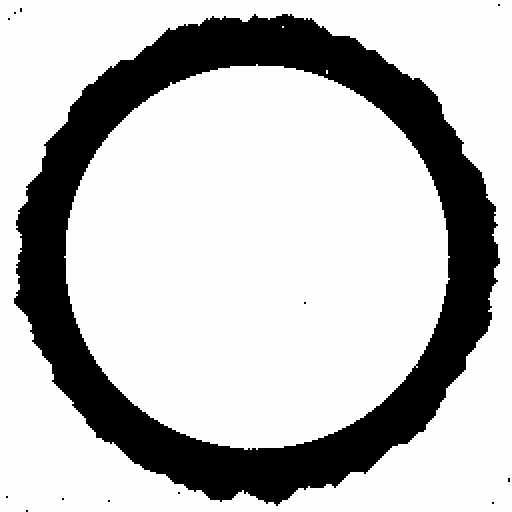}
        \caption{}
        \label{}
    \end{subfigure}
    \begin{subfigure}{0.19\textwidth}
        \includegraphics[width=0.5\linewidth]{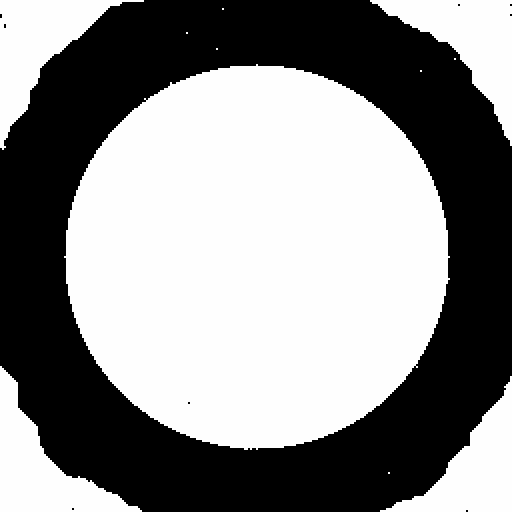}
        \caption{}
        \label{}
    \end{subfigure}
    \begin{subfigure}{0.19\textwidth}
        \includegraphics[width=0.5\linewidth]{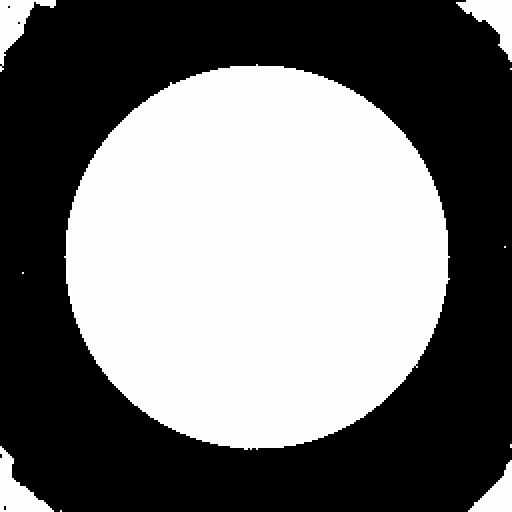}
        \caption{}
        \label{}
    \end{subfigure}

    \begin{subfigure}{0.19\textwidth}
        \includegraphics[width=0.5\linewidth]{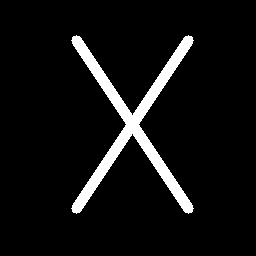}
        \caption{}
        \label{}
    \end{subfigure}%
    \begin{subfigure}{0.19\textwidth}
        \includegraphics[width=0.5\linewidth]{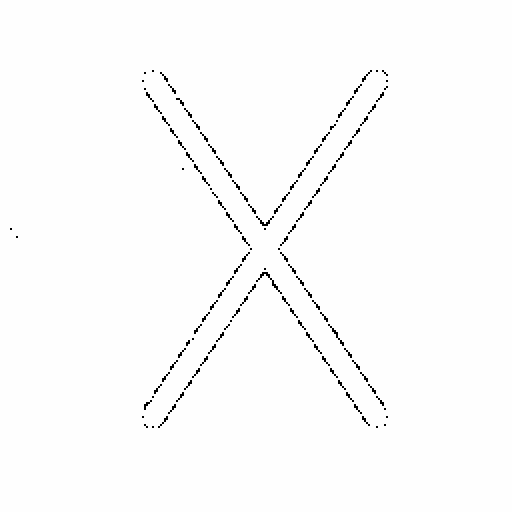}
        \caption{}
        \label{}
    \end{subfigure}
    \begin{subfigure}{0.19\textwidth}
        \includegraphics[width=0.5\linewidth]{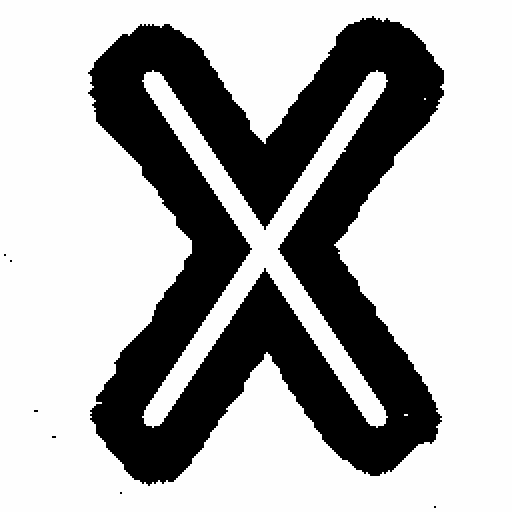}
        \caption{}
        \label{}
    \end{subfigure}
    \begin{subfigure}{0.19\textwidth}
        \includegraphics[width=0.5\linewidth]{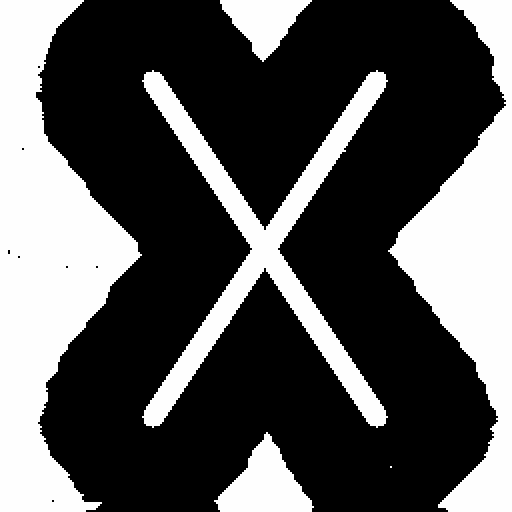}
        \caption{}
        \label{}
    \end{subfigure}
    \begin{subfigure}{0.19\textwidth}
        \includegraphics[width=0.5\linewidth]{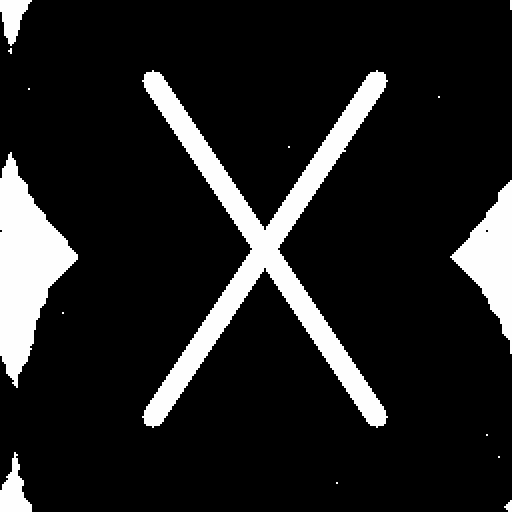}
        \caption{}
        \label{}
    \end{subfigure}

    \begin{subfigure}{0.19\textwidth}
        \includegraphics[width=0.5\linewidth]{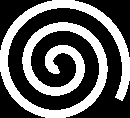}
        \caption{}
        \label{}
    \end{subfigure}%
    \begin{subfigure}{0.19\textwidth}
        \includegraphics[width=0.5\linewidth]{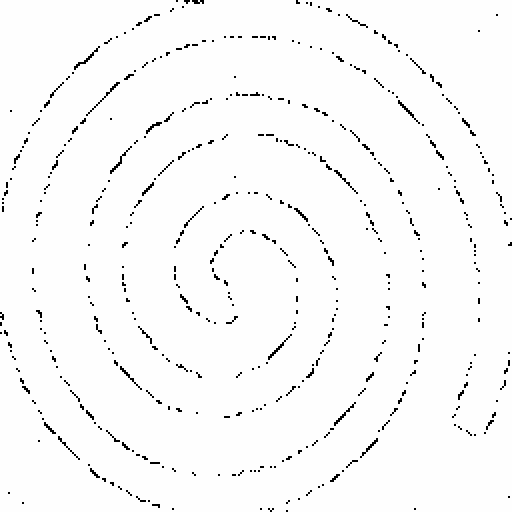}
        \caption{}
        \label{}
    \end{subfigure}
    \begin{subfigure}{0.19\textwidth}
        \includegraphics[width=0.5\linewidth]{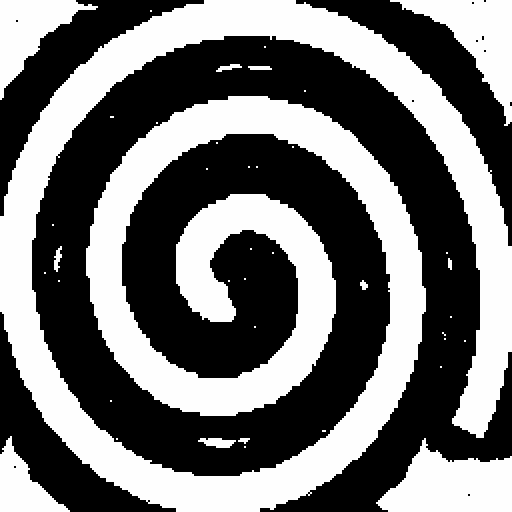}
        \caption{}
        \label{}
    \end{subfigure}
    \begin{subfigure}{0.19\textwidth}
        \includegraphics[width=0.5\linewidth]{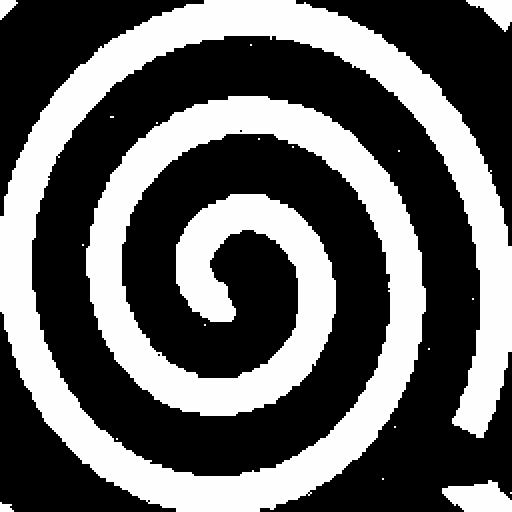}
        \caption{}
        \label{}
    \end{subfigure}
    \begin{subfigure}{0.19\textwidth}
        \includegraphics[width=0.5\linewidth]{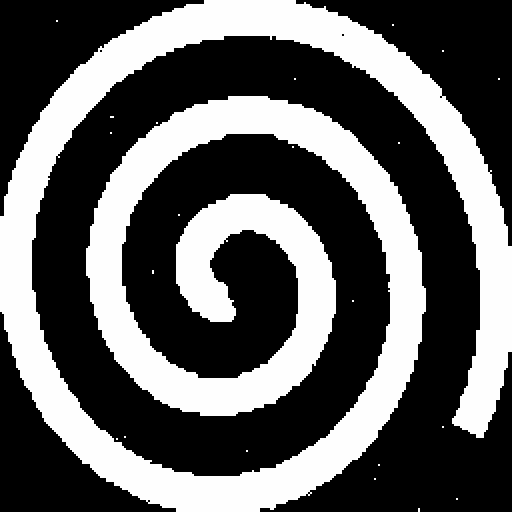}
        \caption{}
        \label{}
    \end{subfigure}    
\caption{Optimization of the exchange parameter $J_{ij}$ between spins in order for a spin distribution to achieve some  pattern, specified in the first column (a,f,k). Each column represent snapshots of the spin lattice at different Monte Carlo steps (every 20 steps).}
\label{fig:spinpattern}
\end{figure}

The results are depicted in Figure \ref{fig:spinpattern}. They not only demonstrate the efficacy of the optimized differentiable Monte Carlo algorithm but also hold implications for manipulation of spin systems. The three distinct patterns achieved in the spin system highlights the adaptability of the algorithm in reaching specific target configurations. This is particularly relevant in the field of material science and condensed matter physics, where the ability to control and predict spin configurations can lead to advancements in magnetic storage technologies and quantum computing.

\subsubsection{Optimization of the spin values}

In this section, we aim to optimize spin values to identify the lowest energy state. We employ this technique on two distinct sets of exchange parameters: firstly, we consider $J>0$, where the lowest energy state corresponds to a ferromagnetic phase; secondly, we examine a set with random J values of -1 or 1, drawn uniformly with a mean of zero, which leads the system to a spin glass phase at low temperatures \cite{parisi_spin_2006}.
The optimization procedure adheres to the algorithm outlined in the Methods section, with the incorporation of an additional step: the quantization of gradients, due to the discrete nature of spin values. Our comparative analysis encompasses three distinct techniques:

\begin{itemize}
    \item Monte Carlo Simulation: Here, we progressively anneal the temperature towards zero concurrent with an increase in the number of MC steps;
    \item Gradient-based Optimization: This technique exclusively relies on updating the spin values according to calculated gradients;
    \item Differentiable Monte Carlo Method: This approach updates spin values by using both MC simulation and gradient-based optimization techniques.
\end{itemize}

\begin{figure}[H]
    \centering
    \begin{subfigure}{0.5\textwidth}
        \includegraphics[width=1\linewidth]{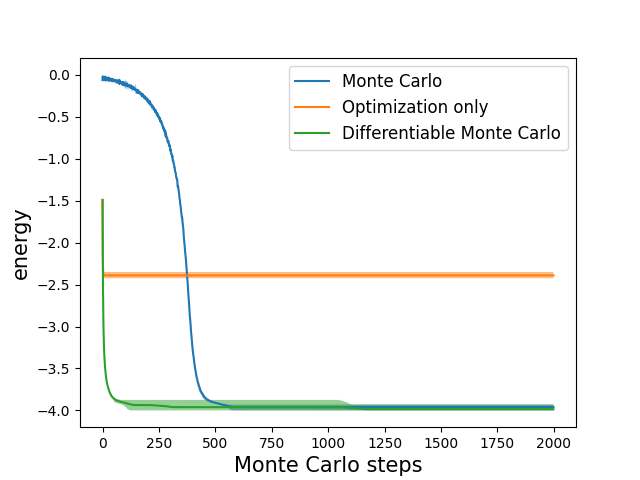}
        \caption{}
        \label{}
    \end{subfigure}%
    \begin{subfigure}{0.5\textwidth}
        \includegraphics[width=1\linewidth]{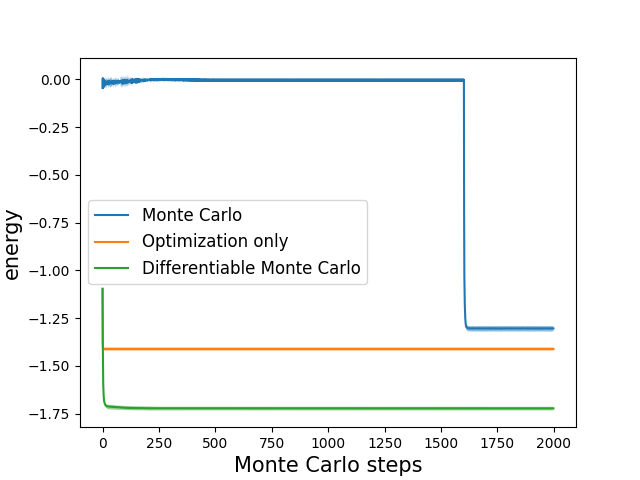}
        \caption{}
        \label{}
    \end{subfigure}
    \caption{ 
    Comparison of Energy Minimization Techniques: Monte Carlo Simulation, Gradient-based Optimization, and Differentiable Monte Carlo. The graph on the left illustrates the scenario with uniform exchange values $J=1$, while the plot on the right depicts random exchange values of $J\in\{-1,1\}$. Notably, Differentiable Monte Carlo consistently achieves the lowest energy state relative to the other techniques. The shaded regions represent the range of energy values encountered at each step, providing insight into the variance between the highest and lowest energies obtained.}
    \label{fig:threemethods}
\end{figure}

Figure \ref{fig:threemethods} presents the system's energy as a function of the number of Monte Carlo steps for the three methods under comparison. The Differentiable Monte Carlo method identifies states of lower energy in contrast to those found by either the Monte Carlo simulation or gradient-based optimization alone. These results imply that DMC effectively leverages gradient information to steer the simulation towards states that minimize the loss function, thereby suggesting a better efficiency in energy landscape navigation.

\section{Conclusions}
\label{sec:conc}

In this paper, we have presented a novel approach for simulating three different spin models using differentiable programming. Our method applies convolution on the spins, similar to how convolution is applied to images in computer vision, which allows us to calculate the Hamiltonian of the system with high accuracy and practicality. In addition, we made use of the checkerboard algorithm to parallelize the calculation of the energy of each spin. This algorithm involves dividing the spins into two sets, with each set updating alternately, such that neighboring spins are updated on different iterations. By doing so, we can parallelize the calculation of the energy of each spin, further improving the efficiency of our approach.

The use of the checkerboard algorithm in conjunction with our approach provides a significant boost in performance, enabling us to simulate spin models with high speed due to the parallelization. We believe that this approach could be widely applicable in many scientific applications that require fast and accurate simulations of complex physical systems.

The use of differentiable programming in this context is particularly useful, as it enables us to leverage the strengths of deep learning techniques for scientific simulations. We demonstrated the effectiveness of our approach by implementing it in PyTorch, which provides easy adaptability to run on GPUs and TPUs. Our experiments show that our method provides a significant speed-up in simulating spin models, without sacrificing accuracy. Moreover, by making use of the batch dimension, we were able to parallelize the simulation even further, leading to an additional increase in performance.

A crucial aspect of our study involved the optimization of spin models, as shown with the optimization of a loss function with respect to the exchange interactions in the Ising model. We observed that our method could effectively adjust parameters to optimize the predefined objective function. This ability to fine-tune parameters in a computationally efficient manner further highlights the potential of our framework in a broader scientific context.

Our work provides a promising direction for future research in this field, as it opens up new opportunities for accelerating simulations and improving our understanding of complex physical phenomena. We anticipate that our method could have a wide range of applications in the future, especially in cases where speed and scalability are essential. By leveraging the power of differentiable programming, we can enable faster simulations and deeper insights into the behavior of physical systems.

\begin{acknowledgments}
This work was supported by the S\~ao Paulo Research Foundation (FAPESP), Grant No. 2023/15739-3, by the National Institute for the Science and Technology of Quantum Information (INCT-IQ), Grant No. 465469/2014-0, and by the National Council for Scientific and Technological Development (CNPq), Grants No. 309862/2021-3, No. 309817/2021-8 and No. 409673/2022-6.
\end{acknowledgments}

\vspace{0.3cm}
\textbf{Data availability.}
The data and code that support the findings of this study are available
at \url{https://github.com/tiago939/dp_monte_carlo}.

\bibliography{bibliography}

\appendix

\end{document}